\documentclass[twocolumn,showpacs,preprintnumbers,amsmath,amssymb]{revtex4}

\usepackage{graphicx}
\usepackage{dcolumn}
\usepackage{bm}
\usepackage{epsfig}
\usepackage{epsfig}
\usepackage{amsmath}
\usepackage{amsfonts}

\newcommand{\ba}{\begin{eqnarray}}
\newcommand{\ea}{\end{eqnarray}}
\newcommand{\be}{\begin{equation}}
\newcommand{\ee}{\end{equation}}
\newcommand{\bea}{\begin{eqnarray}}
\newcommand{\eea}{\end{eqnarray}}

\usepackage{theorem}

\theorembodyfont{\rmfamily}
\newtheorem{remark}{Remark}
\theoremstyle{break}

\def\QED{~\rule[-1pt]{5pt}{5pt}\par\medskip}

\begin{document}


\title{ Controllability of the coupled spin-half harmonic oscillator system}

\author{Haidong Yuan\footnotemark[1],Seth Lloyd}

\affiliation{ Department of Mechanical Engineering, MIT,Cambridge,
MA 02139}

\date{\today}

\begin{abstract}
We present a control-theoretic analysis of the system consisting
of a two-level atom coupled with a quantum harmonic oscillator.
We show that by applying external fields with just two resonant
frequencies, any desired unitary operator can be generated.
\end{abstract}


\maketitle \footnotetext[1]{haidong@mit.edu}
\section{\label{sec:introduction}Introduction}
In this paper, we apply theoretical concepts of quantum control to
the joint system consisting of a two state system coupled with a
quantum harmonic oscillator.  Such systems are ubiquitous in
Nature.  For example, coupled atom-oscillator systems form the
basis for the ion trap quantum computer\cite{CZ95}.  Other
examples include a single atom in a cavity\cite{Kimble}, a
super-conducting qubit in a cavity\cite{Wall}, and control of
single atom lasers\cite{Carm,Mcke}. In \cite{Law}, Law and Eberly
showed that arbitrary states can be synthesized by using just two
resonant frequencies, a result experimentally verified in
\cite{Wineland03}, and \cite{Rangan} showed that the two-level
atom-oscillator system could be controlled by fine-tuning the
Lamb-Dicke parameter.  Here we prove that the dynamics of such
systems is controllable without any fine-tuning or special state
preparation: with the proper sequence of pulses, it is possible to
perform any desired unitary transformation on the Hilbert space
spanned by the atomic states together with the lowest $n$ energy
levels of the oscillator.

In this paper, we will use the ion trap as our model system. An
ion trap quantum computer can be modeled as a collection of $N$
particles with spin $1 \over 2$ in a one-dimensional harmonic
potential. Laser pulses incident on the ions can be tuned to
simultaneously cause internal spin transitions and vibrational
(phonon) excitations, thus allowing local internal states to be
mapped into shared phonon states. The computational qubits are
encoded by two internal states of each ion and the collective
vibration of the trapped ions acts as the information bus. In this
manner, quantum information can be communicated between any pair
of ions and logic gates can be performed. Several key features of
the original proposal in \cite{CZ95}, including the production of
entangled states and the implementation of quantum controlled
operations between a pair of trapped ions, have already been
experimentally demonstrated (see, e.g.,
\cite{Monroe95,King98,Turchette98,Sackett00}).
Meanwhile, several alternative theoretical schemes (see, e.g.,
\cite{Duan01,Jonathan01,wei02,Jonathan00,Childs01,Rangan}) have
also been developed for overcoming various difficulties in
realizing a practical ion-trap quantum information processor. All
these proposals either require fine-tuning of the Lamb-Dicke
parameter or an initial eigenstate of the vibration motion. Here
we present a control theoretical analysis and show that in the
Lamb-Dicke regime by using two resonant frequencies, any unitary
transformation within a finite level of the harmonic oscillator
can be generated.  Unlike, e.g., \cite{Rangan}, no fine-tuning of
the Lamb-Dicke parameter is required to obtain complete control.
While the proof of controllability is somewhat involved, because
of the fundamental nature of the system to be controlled and
because of the wide range of potential application, we present
this proof in detail.  As will be seen below, the difficulty of
the proof arises because, in the absence of controllability of the
Lamb-Dicke parameter, one must combine discrete and continuous
control theoretic techniques. The current proof can be regarded as
extending the techniques of the paper\cite{Childs01,Chuang03} from
controlling 4 states to controlling $m$ states, where $m$ can be
arbitrary large. We begin in section \ref{sec:model} by presenting
the usual Jaynes-Cumming model for spin boson interaction. We then
make the controllability analysis of the system in section
\ref{sec:control}.

\section{Laser-ion interaction model}
\label{sec:model} The physical situation we consider is a
two-state atom (frequency $\omega_c$) coupled to a harmonic
oscillator (transition frequency $\omega_z$), driven additionally
by an external field (frequency $\omega$). We will follow the ion
trap model~\cite{Win97,Childs01}.
The free Hamiltonian of this system is $H_0=\hbar \omega_c
{\sigma_z \over 2} + \hbar \omega_z a^\dagger a$, where $\sigma_z$
is a Pauli spin operator and $a$ annihilates a phonon. Turning on
the electromagnetic field of a laser gives an interaction
Hamiltonian \be \label{eq:interaction} H_I = - \vec\mu \cdot \vec
B \,, \ee where $\vec \mu = \mu \vec \sigma/2$ is the magnetic
moment of the ion and $\vec B=B \hat x \cos(kz-\omega t+\Phi)$ is
the magnetic field produced by the laser.  Here
$z=z_0(a+a^\dagger)$, where $z_0=\sqrt{\hbar/2Nm\omega_z}$ is a
characteristic length scale for the motional wave functions and
$m$ is the mass of an ion.

We consider the regime in which $\eta \equiv k z_0 \ll 1$.  In
this regime, we may determine the effect of a laser pulse at a
specific frequency $\omega$ by expanding
Eq.~(\ref{eq:interaction}) in powers of $\eta$ and neglecting
rapidly rotating terms. Then pulsing on resonance ($\omega =
\omega_c$) allows one to perform the transformation \be
R(\theta,\phi) = \exp\left[i{\theta} ( e^{i\phi} \sigma^+ +
e^{-i\phi} \sigma^- )
      \right]
\,, \ee and pulsing at the red sideband frequency
($\omega=\omega_c-\omega_z$) gives \be \label{eq:red}
R^-(\theta,\phi) = \exp\left[i {\theta} ( e^{i\phi} \sigma^+
a^{\dagger}+ e^{-i\phi} \sigma^- a ) \right] \,. \ee In each case,
the parameter $\theta$ depends on the strength and duration of the
pulse and $\phi$ depends on its phase.
\begin{figure}[h!]
\label{fig:HO} \centering
\includegraphics[width=3in]{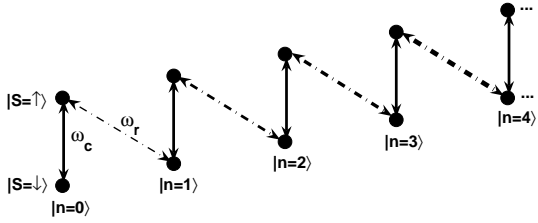}
\caption{Graphical representation of the quantum harmonic
oscillator driven by a sinusoidal resonant field fields $\omega_c$
and $\omega_r=\omega_c-\omega_z$ as shown. The strengths of the
$\omega_c$ transition couplings are independent of the harmonic
oscillator quantum number $n$, whereas the strength of the
$\omega_r$ transition couplings increase as the square root of the
quantum number $n$.}
\end{figure}
In the next section, we show that by just using these two
frequencies any unitary operator can be generated. The basic idea
in proving controllability is an extension of
\cite{Childs01,Chuang03}.  Using the feature that the transition
frequencies increase as the square root of the quantum number, we
apply only pulses that leave the system confined within the
Hilbert space spanned by the first $n$ oscillator levels.  This
requirement means that the only a discrete set of pulses can be
applied at the red sideband frequency.   Meanwhile, a continuous
set of pulses can be applied at the resonance frequency.   As a
result of the use of both discrete and continuous controls, the
resulting control problem is technically somewhat involved.
Nonetheless, it can be solved completely, as we now show.

\section{Controllability analysis}
\label{sec:control} We denote $E_{pq}$ be the matrix such that has
all the entries equal to zero except the $pq$ entry, which equals
$1$. It is easy to check that $E_{pq}E_{rs}=\delta_q^rE_{ps}$.

The Hamiltonian, after absorbing the imaginary number $i$, can be
represented as skew-Hermitian matrices. If we take the eigen-state
of the free Hamiltonian as the basis, then after re-scaling the
time unit, the various Hamiltonians in the interaction frame can
be represented as

1) $\omega=\omega_c$, $\phi=0$ \bea \aligned
H_1 &=i\sum_{k=0}^{\infty}E_{(2k+1)(2k+2)}+E_{(2k+2)(2k+1)}\\
&=i\left[ \begin{array}{ccccccc}
0 & 1 & 0 & 0 & 0 & 0 & ... \\
1 & 0 & 0 & 0 & 0 & 0 & ... \\
0 & 0 & 0 & 1 & 0 & 0 & ... \\
0 & 0 & 1 & 0 & 0 & 0 & ... \\
0 & 0 & 0 & 0 & 0 & 1 & ... \\
0 & 0 & 0 &0 & 1 & 0 & ... \\
\vdots & \vdots & \vdots & \vdots & \vdots & \vdots & \vdots\\
\end {array} \right]
\,,
\endaligned
\eea

2) $\omega=\omega_c$, $\phi=\frac{\pi}{2}$, \bea \aligned
H_2&=\sum_{k=0}^{\infty}E_{(2k+1)(2k+2)}-E_{(2k+2)(2k+1)}\\
&=\left[ \begin{array}{ccccccc}
0 & 1 & 0 & 0 & 0 & 0 & ... \\
-1 & 0 & 0 & 0 & 0 & 0 & ... \\
0 & 0 & 0 & 1 & 0 & 0 & ... \\
0 & 0 & -1 & 0 & 0 & 0 & ... \\
0 & 0 & 0 & 0 & 0 & 1 & ... \\
0 & 0 & 0 &0 & -1 & 0 & ... \\
\vdots & \vdots & \vdots & \vdots & \vdots & \vdots & \vdots\\
\end {array} \right]
\,,
\endaligned
\eea

3) $\omega=\omega_c-\omega_z$, $\phi=0$, \bea \aligned
H_3&=i\sum_{k=1}^{\infty}\sqrt{k}\left[E_{(2k)(2k+1)}+E_{(2k+1)(2k)}\right]
\\&=i\left[ \begin{array}{ccccccc}
0 & 0 & 0 & 0 & 0 & 0 & ... \\
0 & 0 & 1 & 0 & 0 & 0 & ... \\
0 & 1 & 0 & 0 & 0 & 0 & ... \\
0 & 0 & 0 & 0 & \sqrt{2} & 0 & ... \\
0 & 0 & 0 & \sqrt{2} & 0 & 0 & ... \\
0 & 0 & 0 &0 & 0 & 0 & ... \\
\vdots & \vdots & \vdots & \vdots & \vdots & \vdots & \vdots\\
\end {array} \right] .
\endaligned
\eea

4) $\omega=\omega_c-\omega_z$, $\phi=\frac{\pi}{2}$, \bea \aligned
H_4&=\sum_{k=1}^{\infty}\sqrt{k}\left[E_{(2k)(2k+1)}-E_{(2k+1)(2k)}\right]\\
&=\left[ \begin{array}{ccccccc}
0 & 0 & 0 & 0 & 0 & 0 & ... \\
0 & 0 & 1 & 0 & 0 & 0 & ... \\
0 & -1 & 0 & 0 & 0 & 0 & ... \\
0 & 0 & 0 & 0 & \sqrt{2} & 0 & ... \\
0 & 0 & 0 & -\sqrt{2} & 0 & 0 & ... \\
0 & 0 & 0 &0 & 0 & 0 & ... \\
\vdots & \vdots & \vdots & \vdots & \vdots & \vdots & \vdots\\
\end {array} \right] .
\endaligned
\eea

When the Hamiltonian $H_3$ or $H_4$ is applied,
$|\uparrow\rangle|m\rangle$ is connected to
$|\downarrow\rangle|m+1\rangle$. We restrict the evolution time
$T$ under these two Hamiltonian to satisfy $T\sqrt{m}=k\pi$, while
$k$ is integer, so the subspace of states spanned by
$\{|\downarrow,\uparrow\rangle|j\rangle |j\leq m\}$ is preserved.
We show that under these restrictions any unitary matrix within
any finite
harmonic level still can be generated. 
\subsection{$SU(4)$}
Let's first work out the case of $m=1$, show that we can generate
$SU(4)$ on the subspace spanned by states
$$|\downarrow\rangle|0\rangle,|\uparrow\rangle|0\rangle,|\downarrow\rangle|1\rangle,|\uparrow\rangle|1\rangle$$
Restricted to this subspace,
$$H_1 = i\left[
\begin{array}{cccc}
0 & 1 & 0 & 0  \\
1 & 0 & 0 & 0  \\
0 & 0 & 0 & 1  \\
0 & 0 & 1 & 0  \\
\end {array} \right]
\,,$$

$$H_2 = \left[
\begin{array}{cccc}
0 & 1 & 0 & 0  \\
-1 & 0 & 0 & 0  \\
0 & 0 & 0 & 1  \\
0 & 0 & -1 & 0  \\
\end {array} \right]
\,,$$ and the unitary operators we can generate using $H_3$ and
$H_4$ are \bea \aligned
R^-(\frac{k\pi}{\sqrt{2}},0)&=\exp\left[\frac{k\pi}{\sqrt{2}}H_3\right]\\
&=\left[\begin{array}{cccc}
1 & 0 & 0 & 0  \\
0 & \cos(\frac{k\pi}{\sqrt{2}}) & i\sin(\frac{k\pi}{\sqrt{2}}) & 0  \\
0 & i\sin(\frac{k\pi}{\sqrt{2}}) & \cos(\frac{k\pi}{\sqrt{2}})0 & 0  \\
0 & 0 & 0 & (-1)^k  \\
\end {array} \right]
\,,
\endaligned
\eea

\bea \aligned
R^-(\frac{k\pi}{\sqrt{2}},\frac{\pi}{2})&=\exp\left[\frac{k\pi}{\sqrt{2}}H_4\right]\\
&=\left[\begin{array}{cccc}
1 & 0 & 0 & 0  \\
0 & \cos(\frac{k\pi}{\sqrt{2}}) & \sin(\frac{k\pi}{\sqrt{2}}) & 0  \\
0 & -\sin(\frac{k\pi}{\sqrt{2}}) & \cos(\frac{k\pi}{\sqrt{2}}) & 0  \\
0 & 0 & 0 & (-1)^k  \\
\end {array} \right]
\,,
\endaligned
\eea Choose $k=2p$, by varying $p$,
$R^-(\frac{k\pi}{\sqrt{2}},\frac{\pi}{2})$ forms a dense subset of
the one parameter group \be \nonumber e^{t \left[
\begin{array}{cccc}
0 & 0 & 0 & 0  \\
0 & 0 & 1 & 0  \\
0 & -1 & 0 & 0  \\
0 & 0 & 0 & 0  \\
\end {array} \right]}
\ee Thus we have the generator $\left[
\begin{array}{cccc}
0 & 0 & 0 & 0  \\
0 & 0 & 1 & 0  \\
0 & -1 & 0 & 0  \\
0 & 0 & 0 & 0  \\
\end {array} \right]$, add it to $H_2$, we get
$$H_5=\left[
\begin{array}{cccc}
0 & 1 & 0 & 0  \\
-1 & 0 & 1 & 0  \\
0 & -1 & 0 & 1  \\
0 & 0 & -1 & 0  \\
\end {array} \right]$$
Choose $k=2p+1$, we can have \bea \aligned
U_1&=R^-(\frac{k\pi}{\sqrt{2}},\frac{\pi}{2})\\
&=\left[\begin{array}{cccc}
1 & 0 & 0 & 0  \\
0 & \cos(\frac{k\pi}{\sqrt{2}}) & \sin(\frac{k\pi}{\sqrt{2}}) & 0  \\
0 & -\sin(\frac{k\pi}{\sqrt{2}}) & \cos(\frac{k\pi}{\sqrt{2}}) & 0  \\
0 & 0 & 0 & -1  \\
\end {array} \right]
\,,
\endaligned
\eea Since $U(4)$ is compact, the infinite sequence
$\{U_1,U_1^2,U_1^3,U_1^4...\}$ has a convergent subsequence, i.e.,
there exists $p_1>p_2\in N$ such that $U_1^{p_1}-U_1^{p_2}$ is
arbitrary close to zero, when this is true then $U_1^{p_1-p_2-1}$
is arbitrary close
to $U_1^{-1}$, i.e., $U_1^{-1}$ can be approximately generated to arbitrary accuracy. But \bea \aligned U_1^{-1}&H_1U_1\\
&= i\left[\begin{array}{cccc}

0 & \cos(\frac{k\pi}{\sqrt{2}}) & \sin(\frac{k\pi}{\sqrt{2}}) & 0  \\
\cos(\frac{k\pi}{\sqrt{2}}) & 0 & 0 & \sin(\frac{k\pi}{\sqrt{2}})  \\
\sin(\frac{k\pi}{\sqrt{2}}) & 0 & 0 & -\cos(\frac{k\pi}{\sqrt{2}})   \\
0 & \sin(\frac{k\pi}{\sqrt{2}}) & -\cos(\frac{k\pi}{\sqrt{2}}) & 0\\

\end {array} \right]
\,,
\endaligned
\eea choose $k$ such that $\frac{k}{\sqrt{2}}$ is arbitrary close
to an integer, we can get the transformation
$$i\left[
\begin{array}{cccc}
0 & 1 & 0 & 0  \\
1 & 0 & 0 & 0  \\
0 & 0 & 0 & -1  \\
0 & 0 & -1 & 0  \\
\end {array} \right]$$
Subtracting this from $H_1$ and dividing by a factor $2$ yields
$$H_6=i\left[
\begin{array}{cccc}
0 & 0 & 0 & 0  \\
0 & 0 & 0 & 0  \\
0 & 0 & 0 & 1  \\
0 & 0 & 1 & 0  \\
\end {array} \right]$$
Similarly by using $H_2$ and $U_1$, we can get $$H_7=\left[
\begin{array}{cccc}
0 & 0 & 0 & 0  \\
0 & 0 & 0 & 0  \\
0 & 0 & 0 & 1  \\
0 & 0 & -1 & 0  \\
\end {array} \right]$$
We now show that $\{H_5,H_6,H_7\}$ generate all the skew-Hermitian
matrices on the subspace. First
$$H_8=H_5-H_7=\sum_{k=1}^{N-2}E_{k(k+1)}-E_{(k+1)k}$$
Here $N=4$, we will do the following computation using the general
$N$, as this will be used for the proof of the general case. Now,
$$H_7=E_{(N-1)N}-E_{N(N-1)}$$
We first show that $H_7$ and $H_8$ generate all the real
skew-symmetric matrices of size $N\times
N$\cite{Brockett,Sussman}, let \bea \aligned \nonumber
M_{N-1}&=H_7=E_{(N-1)N}-E_{N(N-1)}\\
M_{N-2}=&[H_8,M_{N-1}]=E_{(N-2)N}-E_{N(N-2)}\\
M_{N-3}=&[H_8,M_{N-2}]+M_{N-1}=E_{(N-3)N}-E_{N(N-3)}\\
M_{N-4}=&[H_8,M_{N-3}]+M_{N-2}=E_{(N-4)N}-E_{N(N-4)}\\
   &\vdots\\
M_{1}=&[H_8,M_{2}]+M_3=E_{1N}-E_{N1}\\
\endaligned
\eea and $[M_p,M_q]=E_{qp}-E_{pq}$, $\forall p\neq q\in
\{1,2,...,N-1\}$. Thus we can generate complete basis for
skew-symmetric matrices. Similarly \bea \aligned \nonumber
J_{N-1}&=H_6=i(E_{(N-1)N}+E_{N(N-1)})\\
J_{N-2}=&[H_8,J_{N-1}]=i(E_{(N-2)N}+E_{N(N-2)})\\
J_{N-3}=&[H_8,J_{N-2}]+J_{N-1}=i(E_{(N-3)N}+E_{N(N-3)})\\
   &\vdots\\
J_{1}=&[H_8,J_{2}]+J_{3}=i(E_{1N}+E_{N1})\\
\endaligned
\eea and $$[M_q,J_p]=i(E_{qp}+E_{pq})$$
$$[i(E_{qp}+E_{pq}),E_{qp}-E_{pq}]=2i(E_{pp}-E_{qq})$$
$\forall p\neq q\in \{1,2,...,N-1\}$. So we can generate a full
basis for all $N\times N$ skew-Hermitian matrices. This proves the $SU(4)$ case.

\subsection{General case}
Now we generalize our proof to the controllability on $SU(n)$ for
any $n$. It is not necessary to check the case for each $n$, as
$SU(n_1)$ is a subgroup of $SU(n_2)$, for $n_1<n_2$, the
controllability on $SU(n_2)$ implies controllability on $SU(n_1)$.
It is sufficient to prove the result for infinitely many $n_i$ as
$n_i\rightarrow\infty$.

Take the subspace up to Harmonic level $m$, i.e.,
$$\{|\downarrow\rangle|0\rangle,|\uparrow\rangle|0\rangle,|\downarrow\rangle|1\rangle,|\uparrow\rangle|1\rangle,...,|\downarrow\rangle|m\rangle,|\uparrow\rangle|m\rangle\}$$
where $(m-1,m+1)$ are both prime. We shall prove the
controllability on this subspace. The twin prime conjecture claims
there exists infinitely many such primes. If the twin prime
conjecture is false, then the following proof works only up to
$n=2m+2$, where $m$ is the largest known twin prime. As of $2006$,
the largest known twin prime is $100314512544015\cdot2^{171960}\pm
1$, which is large enough for most physical systems. Below, we
generalize the twin-prime proof to show controllability for all
$n$.

If we restrict the evolution time $T$ for $H_4$ to satisfy
$T\sqrt{m+1}=k\pi$, where $k$ is an integer, then the angle
rotated between $|\uparrow\rangle|p-1\rangle$ and
$|\downarrow\rangle|p\rangle$ is
$\sqrt{p}T=k\sqrt{\frac{p}{m+1}}\pi$. We divide the numbers
$\{1,2,...,m\}$ into groups $G_i, i=1,2,...$, such that in same
group $G_i$ the angles rotated under the above evolution are
rationally related to each other, i.e., $p_1$, $p_2$ are in same
group, if and only if $\sqrt{\frac{p_1}{p_2}}$ is a rational
number. For example, $\{1,1\cdot2^2,1\cdot3^2,...,1\cdot p_1^2\}$
forms a group, where $p_1^2\leq m$, $(p_1+1)^2>m$, similarly other
groups are $\{2,2\cdot2^2,2\cdot3^2,...,2\cdot
p_2^2\}$,$\{3,3\cdot2^2,3\cdot3^2,...,3\cdot p_2^2\}$...,
specially $\{m-1\}$ itself forms a group.

As $m+1$ is a prime number, $k\sqrt{\frac{p}{m+1}}(mod 2)$ are
irrational numbers for all $p\leq m$. Accordingly, we can vary $k$
such that, except the angles relate to one group $G_i$, all the
other angles are arbitrary close to zero. This way we can
construct the generator
$$\hat{H}_i=\sum_{j\in G_i}\sqrt{j}(E_{2j(2j+1)}-E_{(2j+1)2j})$$
Add all $\hat{H}_i$ to $H_2$, we get a matrix similar to $H_5$ in
the $SU(4)$ section, with only nonzero entries at the first
off-diagonal. Denote this matrix by $\tilde{H}_5$.

To prove the controllability, we just need to show that we can
also generate matrices similar to $H_6$ and $H_7$, i.e.,
$$E_{(N-1)N}-E_{N(N-1)}$$ and
      $$i(E_{(N-1)N}+E_{N(N-1)})$$
      here $N=2m+2$.

As $\{m-1\}$ itself forms a group, say $G_j$, we can generate
$$S_1=\frac{1}{\sqrt{m-1}}\hat{H}_j=E_{(2m-2)(2m-1)}-E_{(2m-1)(2m-2)}$$
bracket it with
$H_2=\sum_{k=0}^{m}E_{(2k+1)(2k+2)}-E_{(2k+2)(2k+1)}$, we can get
\bea \aligned \nonumber
S_2=&[H_2,S_1]\\
=&E_{(2m-3)(2m-1)}-E_{(2m-1)(2m-3)}\\
&+E_{2m(2m-2)}-E_{(2m-2)2m}
\endaligned
\eea then bracket $S_2$ with $S_1$, \bea \aligned \nonumber
S_3&=[S_1,S_2]\\
&=E_{(2m-3)(2m-2)}-E_{(2m-2)(2m-3)}\\
&+E_{(2m-1)2m}-E_{2m(2m-1)}
\endaligned
\eea we see that $S_3$ is nothing but the restriction of $H_2$ on
the subspace spanned by
$$|\downarrow\rangle|m-2\rangle,|\uparrow\rangle|m-2\rangle,|\downarrow\rangle|m-1\rangle,|\uparrow\rangle|m-1\rangle$$
Similarly $H_1$ can also be restricted to this subspace. From the
$SU(4)$ case, we know we can generate any skew-Hermitian matrix on
this subspace, specifically we can have
$S_4=E_{(2m-1)2m}-E_{2m(2m-1)}$

Now pick the group $G_p$ to which $m$ belongs. We get \bea
\aligned
 \hat{H}_p=&\sum_{j\in G_p}\sqrt{j}(E_{2j(2j+1)}-E_{(2j+1)2j})\\
          =&E_{2m(2m+1)}-E_{(2m+1)2m}\\
          &+\sum_{j\neq m\in G_p}\sqrt{j}(E_{2j(2j+1)}-E_{(2j+1)2j})
\endaligned
\eea Bracket $S_4$ with $\hat{H}_p$, since all the numbers in
$G_p$ have the form $\frac{m}{q^2}$, the second term in the right
side of the above equation commute with $S_4$. Accordingly we
obtain
$$S_5=[S_4,\hat{H}_p]=E_{(2m-1)(2m+1)}-E_{(2m+1)(2m-1)}$$
Now, bracket $S_5$ with $S_4$,
$$S_6=[S_5,S_4]=E_{2m(2m+1)}-E_{(2m+1)2m}$$
comparing with $S_1$, we see that we just moved one block down.
Repeat what we did with $S_1$ to $S_6$, we can get
 \bea \aligned
 S_7&=E_{(2m+1)(2m+2)}-E_{(2m+2)(2m-1)}\\
   &=E_{(N-1)N}-E_{N(N-1)}
\endaligned
\eea This is the matrix we need to generalize our proof of
controllability on $SU(4)$. Similarly, we can get
$$i(E_{(N-1)N}+E_{N(N-1)}),$$ together with $\tilde{H}_5$, we are
able to generate all the skew-Hermitian matrices of size $N\times
N$, which proves the controllability on $SU(N)$. This completes
the proof: driving the fundamental frequency and the red sideband
suffice to control the two-level atom coupled to an harmonic
oscillator.

\begin{remark}{\rm From the proof we see that the only two properties of
the
pair $(m-1,m+1)$ we used are: \\
1: $\sqrt{\frac{p}{m+1}}$ are irrational for all $p\leq m$. \\
2: There exists one group consists of only one number.\\
It is convenient to pick twin primes, but there exist other
choices. For example, we can choose $m+1=2q$, where $q$ is an odd
prime. Under this choice condition 1 still holds, and $q$ itself
forms a group. So our proof, while expressed in terms of the twin
prime conjecture, actually holds for all $n$. }
\end{remark}

\section{Discussion}

We have proved the controllability of the dynamics of the coupled
two-level system/harmonic oscillator.  Because of the discrete
nature of the controls, the proof was somewhat involved.  In
addition, the system is only fully controllable in the limit that
the number of control pulses goes to infinity.    In any realistic
setting we will have only a finite time and a finite number of
pulses that we can apply.   The question of the rate of
convergence of such discrete schemes is an important open question
in control theory and in quantum information, and will be
investigated elsewhere. For the moment, we note only that
accurately generating arbitary members of $SU(4)$ and $SU(n)$ for
$n\leq 10$ or so via the techniques described here is well within
the reach of current experiment.


\end{document}